# Univariate polynomial real root isolation:
# Continued Fractions revisited

Elias P. Tsigaridas and Ioannis Z. Emiris

Department of Informatics and Telecommunications

National Kapodistrian University of Athens, HELLAS

{et,emiris}@di.uoa.gr

September 6, 2018

## Abstract

We present algorithmic, complexity and implementation results concerning real root isolation of integer univariate polynomials using the continued fraction expansion of real algebraic numbers. One motivation is to explain the method's good performance in practice. We improve the previously known bound by a factor of $d\tau$, where $d$ is the polynomial degree and $\tau$ bounds the coefficient bitsize, thus matching the current record complexity for real root isolation by exact methods. Namely, the complexity bound is $\widetilde{\mathcal{O}}_B(d^4\tau^2)$ using the standard bound on the expected bitsize of the integers in the continued fraction expansion. We show how to compute the multiplicities within the same complexity and extend the algorithm to non square-free polynomials. Finally, we present an efficient open-source C++ implementation in the algebraic library SYNAPS, and illustrate its efficiency as compared to other available software. We use polynomials with coefficient bitsize up to 8000 and degree up to 1000.

## 1 Introduction

In this paper we deal with real root isolation of univariate integer polynomials, a fundamental problem in computer algebra as well as in many applications ranging from computational geometry to quantifier elimination. The problem consists in computing intervals with rational endpoints which contain exactly one real root of the polynomial. We use the continued fraction expansion of *real algebraic numbers*. Recall that such a number is a real root of an integer polynomial.

Another motivation is to explain the method's good performance in implementations, albeit the higher complexity bounds which was known until now. Indeed, we show that continued fractions lead to asymptotic bit complexity bounds that match those recently proven for other exact methods, such as Sturm sequences and Descartes' subdivision.

### 1.1 Notation

In what follows $\mathcal{O}_B$ means bit complexity and the $\widetilde{\mathcal{O}}_B$-notation means that we are ignoring logarithmic factors. For a polynomial $A = \sum_{i=1}^{d} a_i X^i \in \mathbb{Z}[X]$, $\deg(A)$ denotes its degree. We consider square-free polynomials except if explicitly stated otherwise. By $\mathcal{L}(A)$ we denote an upper bound on the bit size of the coefficients of $A$ (including a bit for the sign). For $\mathtt{a} \in \mathbb{Q}$, $\mathcal{L}(\mathtt{a}) \geq 1$ is the maximum bit size of the numerator and the denominator. Let $\mathsf{M}(\tau)$ denote the bit complexity of multiplying two integers of bit size at most $\tau$. Using FFT, $\mathsf{M}(\tau) = \mathcal{O}_B(\tau \lg^c \tau)$ for a suitable constant $c$. $Var(A)$ denotes the sign variations in the coefficient list of $A$ ignoring zero terms and $\Delta$ the separation bound of $A$, that is the smallest distance between two (complex) roots of $A$. Finally $N = \max\{d, \tau\}$.

## 1.2 Previous work and our results

Real root isolation of univariate integer polynomials is a well known problem and various algorithms exist for it. Moreover, there is a huge bibliography on the problem so we have to mention that we only scratch the surface of the existing literature and we encourage the reader to refer to the references.

Exact subdivision based algorithms for real root isolation are based either on Descartes' rule of sign or on Sturm sequences. Roughly speaking, the idea behind both approaches is to subdivide a given interval that initially contains all the real roots until it is certified that none or one root is contained. Descartes' approach achieves this by repeatedly transforming the original polynomial and counting the sign variations in the coefficients' list, while Sturm's approach computes a signed polynomial remainder sequence and evaluates it over the endpoints of the interval of interest. Quite recently it was proven (cf [16, 17] and references therein) that both approaches, the one based on Descartes' rule of sign (where the polynomials are represented either in the monomial or in the Bernstein basis) and the one based on Sturm sequences achieve the same bit complexity bound, namely $\widetilde{\mathcal{O}}_B(d^4\tau^2)$ or $\widetilde{\mathcal{O}}_B(N^6)$. Moreover using Sturm sequences in a pre-processing and a post-processing step [21] the bound holds for the non square-free case and the multiplicities of the roots can also be computed.

The continued fraction algorithm (from now on called CF) differs from the subdivision based algorithms in that instead of bisecting a given initial interval it computes the continued fraction expansions of the real roots of the polynomial. The first formulation of the algorithm is due to Vincent [40], see also [2] for historical references, based on his theorem (Th. 4 without the terminating condition) where it was stated that repeated transformations of the polynomial will eventually yield a polynomial with zero (or one) sign variation, thus Descartes' rule implies the transformed polynomial has zero (resp. one) real root in $(0, \infty)$. If one sign variation is attained then the inverse transformation can be applied in order to compute an isolating interval for the real root that corresponds to the original polynomial and moreover the $c_i$'s appear in the transformation correspond to the partial quotients of the continued fraction expansion of the real root. However Vincent's algorithm is exponential [13]. He computed the $c_i$'s in the transformation of Th. 4 by repeated shift operations of the form $X \mapsto X + 1$, thus if one of the $c_i$'s (or even the sum of all) is of magnitude, say, $2^\tau$ then an exponential number of steps must be performed.

Uspensky [37] extended Vincent's theorem by computing an upper bound on the number of transformations so as to isolate the real roots, but failed to deal with its exponential behavior. See also [12, 32] where the problem of approximating a real algebraic number is also considered. Using Vincent's theorem, Collins and Akritas [13] derived a polynomial subdivision-based algorithm using Descartes' rule of sign. Akritas [3, 1] dealt with the exponential behavior of the CF algorithm, by computing the $c_i$'s in the transformations as positive lower bounds of the positive real roots, via Cauchy's bound (for details, see sec. 3). He achieved a complexity of $\widetilde{\mathcal{O}}_B(d^5\tau^3)$ or $\widetilde{\mathcal{O}}_B(N^8)$, without using fast Taylor shifts [41]. However, it is not clear how this approach accounts for the increased coefficient size in the transformed polynomial after applying $X \mapsto b + X$. Another issue is to bound the size of the $c_i$. Refer to Eq. (1) which indicates that the *magnitude* of the partial quotients is unbounded. CF is the standard real root isolation function in MATHEMATICA [4] and for some experiments against subdivision-based algorithms, also in MATHEMATICA, the reader may refer to [5].

Another class of univariate solvers are numerical solvers, e.g. [30, 8, 9] that compute an approximation of all the roots (real and complex) of a polynomial up to a desired accuracy.

The contributions of this paper are the following: First, we improve the bound of the number of steps (transformations) that the algorithm performs. This is basically achieved through Th. 6. Second, we bound the bitsize of the partial quotients and thus the growth of the transformed polynomials which appear during the algorithm. For this we use the theory of the continued fraction expansion of real numbers and a standard worst case analysis. We revisit the proof of [3, 1] so as to improve the overall bit complexity bound of the algorithm to $\widetilde{\mathcal{O}}_B(N^6)$, thus matching the current record complexity for real root isolation. The extension to the non square-free case uses the techniques from [18]. Third, we present our efficient open-source C++ implementation and illustrate it on various data sets, including polynomials of degree up to 1000 and coefficients of 8000 bits. Our software seems faster than the root-isolation implementations that we tested, including RS. We also tested a numeric solver, namely ABERTH, which has comparable efficiency and, on many instances, is slower. We believe that our software contributes towards reducing the gap between rational and numeric computation, the latter being usually perceived as faster.

The rest of the paper is structured as follows. The next section sketches the theory behind continued fractions. Sec. 3 presents the CF algorithm and Sec. 4 its analysis. We conclude with experiments using our implementation, along with comparisons against other available software for univariate equation solving.

## 2 Continued fractions

We present a short introduction to continued fractions, following [38] which although is far from complete suffices for our purposes. The reader may refer to e.g [3, 42, 10, 38]. In general a *simple (regular) continued fraction* is a (possibly infinite) expression of the form

$$c_0 + \cfrac{1}{c_1 + \cfrac{1}{c_2 + \dots}} = [c_0, c_1, c_2, \dots]$$

where the numbers $c_i$ are called *partial quotients*, $c_i \in \mathbb{Z}$ and $c_i \geq 1$ for $i > 0$. Notice that $c_0$ may have any sign, however in our real root isolation algorithm $c_0 \geq 0$. By considering the recurrent relations

$$P_{-1} = 1, \quad P_0 = c_0, \quad P_{n+1} = c_{n+1} P_n + P_{n-1}$$
$$Q_{-1} = 0, \quad Q_0 = 1, \quad Q_{n+1} = c_{n+1} Q_n + Q_{n-1}$$

it can be shown by induction that $R_n = \frac{P_n}{Q_n} = [c_0, c_1, \dots, c_n]$, for $n = 0, 1, 2, \dots$ and moreover that

$$P_n Q_{n+1} - P_{n+1} Q_n = (-1)^{n+1}$$
$$P_n Q_{n+2} - P_{n+2} Q_n = (-1)^{n+1} c_{n+2}$$

If $\gamma = [c_0, c_1, \dots]$ then $\gamma = c_0 + \frac{1}{Q_0 Q_1} - \frac{1}{Q_1 Q_2} + \dots = c_0 + \sum_{n=1}^{\infty} \frac{(-1)^{n-1}}{Q_{n-1} Q_n}$ and since this is a series of decreasing alternating terms it converges to some real number $\gamma$. A finite section $R_n = \frac{P_n}{Q_n} = [c_0, c_1, \dots, c_n]$ is called the $n$−th *convergent* (or *approximant*) of $\gamma$ and the tails $\gamma_{n+1} = [c_{n+1}, c_{n+2}, \dots]$ are known as its *complete quotients*. That is $\gamma = [c_0, c_1, \dots, c_n, \gamma_{n+1}]$ for $n = 0, 1, 2, \dots$. There is a one to one correspondence between the real numbers and the continued fractions, where evidently the finite continued fractions correspond to rational numbers.

It is known that $Q_n \geq F_{n+1}$ and that $F_{n+1} < \phi^n < F_{n+2}$, where $F_n$ is the $n$−th Fibonacci number and $\phi = \frac{1+\sqrt{5}}{2}$ is the *golden ratio*. Continued fractions are the best (for a given denominator size), approximations, i.e

$$\frac{1}{Q_n(Q_{n+1} + Q_n)} \leq \left| \gamma - \frac{P_n}{Q_n} \right| \leq \frac{1}{Q_n Q_{n+1}} \leq \frac{1}{Q_n^2} < \phi^{-2n}$$

Let $\gamma = [c_0, c_1, \dots]$ be the continued fraction expansion of a real number. The Gauss-Kuzmin distribution [10, 31] states that for almost all real numbers $\gamma$ (meaning that the set of exceptions has measure zero) the probability for a positive integer $\delta$ to appear as an element in the continued fraction expansion of $\gamma$ is

$$Prob[c_i = \delta] = \lg \frac{(\delta + 1)^2}{\delta(\delta + 2)}, \quad i > 0 \tag{1}$$

The Gauss-Kuzmin law induces that we can not bound the mean value of the partial quotients or in other words that the expected value (arithmetic mean) of the partial quotients is diverging, $E[c_i] = \sum_{\delta=1}^{\infty} \delta \, Prob[c_i = \delta] = \infty$, $i > 0$. Surprisingly enough the geometric (and the harmonic)

---

**Algorithm 1:** CF$(A, M)$

---

**Input**: $A \in \mathbb{Z}[X], M(X) = \frac{kX+l}{mX+n}, k, l, m, n \in \mathbb{Z}$
**Output**: A list of isolating intervals

**1** **if** $A(0) = 0$ **then**
**2**      OUTPUT Interval( $M(0), M(0)$ ) ;
**3**      $A \leftarrow A(X)/X$;
**4**      CF$(A, M)$;

**5** $V \leftarrow$ Var$(A)$;
**6** **if** $V = 0$ **then** RETURN ;
**7** **if** $V = 1$ **then**
**8**      OUTPUT Interval( $M(0), M(\infty)$);
**9**      RETURN ;

**10** $b \leftarrow$ PLB$(A)$ **// PLB $\equiv$ PositiveLowerBound** ;
**11** **if** $b > 1$ **then** $A \leftarrow A(b+X), M \leftarrow M(b+X)$ ;

**12** $A_1 \leftarrow A(1+X), M_1 \leftarrow M(1+X)$ ;
**13** CF$(A_1, M_1)$ **// Looking for real roots in $(1, +\infty)$**;
**14** $A_2 \leftarrow A(\frac{1}{1+X}), M_2 \leftarrow M(\frac{1}{1+X})$ ;
**15** CF$(A_2, M_2)$ **// Looking for real roots in $(0, 1)$** ;
**16** RETURN ;

---

mean is not only asymptotically bounded, but is bounded by a constant. For the geometric mean this is the famous Khintchine's constant [23], i.e.

$$\lim_{n \to \infty} \sqrt[n]{\prod_{i=1}^{n} c_i} = \mathcal{K} = 2.685452001...$$

which is an irrational number. The reader may refer to [6] for a comprehensive treatment of *Khintchine's means*. The expected value of the bitsize of the partial quotients is a constant for almost all real numbers, when $n \to \infty$ or $n$ sufficiently big [23, 31]. Following closely [31], we have: $E[\ln c_i] = \frac{1}{n} \sum_{i=1}^{n} \ln c_i = \ln \mathcal{K} = 0.98785...$, as $n \to \infty$, $\forall i > 0$. Let $\mathcal{L}(c_i) \triangleq b_i$, then

$$E[b_i] = \mathcal{O}(1) \tag{2}$$

A real number has an (eventually) periodic continued fraction expansion if and only if it is a root of an irreducible quadratic polynomial. "There is no reason to believe that the continued fraction expansions of non-quadratic algebraic irrationals generally do anything other than faithfully follow Khintchine's law" [11], and also various experimental results [10, 31, 32] suggest so. For the largest digit that can appear in the partial quotients of a rational number the reader may refer to [22].

## 3 The CF algorithm

**Theorem 1 (Descartes' rule of sign)** *The number $R$ of real roots of $A(X)$ in $(0, \infty)$ is bounded by $Var(A)$ and we have $R \equiv Var(A) \mod 2$.*

**Remark 2** *In general Descartes' rule of sign obtains an overestimation of the number of the positive real roots. However if we know that $A$ is hyperbolic, i.e has only real roots or when the number of sign variations is 0 or 1 then it counts exactly.*

**Theorem 3 (Budan)** *[26, 3] Let a polynomial $A$, such that $\deg(A) = d$ and let $a < b$, where $a, b \in \mathbb{R}$. Let $A_a$, resp. $A_b$, be the polynomial produced after we apply the map $X \mapsto X + a$, resp.*

$X \mapsto X + b$, to $A$. Then the followings hold: (i) $Var(A_a) \geq Var(A_b)$, (ii) $\#\{\gamma \in (a,b)|A(\gamma) = 0\} \leq Var(A_a) - Var(A_b)$ and (iii) $\#\{\gamma \in (a,b)|A(\gamma) = 0\} \equiv Var(A_a) - Var(A_b) \mod 2$.

The CF algorithm depends on the following theorem, which dates back to Vincent's theorem in 1836 [40]. The inverse of Th. 4 that proves the termination of CF can be found in [3, 14, 26]. It is a very interesting question whether the one and two circle theorems (c.f [25] and references therein), employed in the analysis of the subdivision-based real-root isolation algorithm [13], can also be applied and possibly improve the complexity of the CF algorithm.

**Theorem 4** *[3, 37] Let $A \in \mathbb{Z}[X]$, with $\deg(A) = d$ and let $\Delta > 0$ be the separation bound. Let $n$ be the smallest index such that*

$$F_{n-1}\Delta > 2 \quad and \quad F_{n-1}F_n\Delta > 1 + \frac{1}{\epsilon_d}$$

*where $F_n$ is the n-th Fibonnaci number and $\epsilon_d = (1 + \frac{1}{d})^{\frac{1}{d-1}} - 1$. Then the map $X \mapsto [c_0, c_1, \ldots, c_n, X]$, where $c_0, c_1, \ldots, c_n$ is an arbitrary sequence of positive integers, transforms $A(X)$ to $A_n(X)$, which has no more than one sign variation.*

**Remark 5** *Since $\frac{3}{4d^2} < \epsilon_d < \frac{4}{d^2}$ [14] we conclude that $\frac{1}{\epsilon_d} + 1 < 2d^2$ for $d \geq 2$. Thus, if $d \geq 2$ we can replace the two conditions of Th. 4 by $F_{n-1}\Delta \geq 2d^2$, since $F_n \geq F_{n-1} \geq 1$ and $F_{n-1}F_n\Delta \geq F_{n-1}\Delta \geq 2d^2 > 2$.*

Th. 4 can be used to isolate the positive real roots of a square-free polynomial $A$. In order to isolate the negative roots we perform the transformation $X \mapsto -X$, so in what follows we will consider only the positive real roots of $A$.

Vincent's variant of the CF algorithm goes as follows: A polynomial $A$ is transformed to $A_1$ by the transformation $X \mapsto 1 + X$ and if $Var(A_1) = 0$ or $Var(A_1) = 1$ then $A$ has 0, resp. 1, real root greater than 1 (Th. 1). If $Var(A_1) < Var(A)$ then (possibly) there are real roots of $A$ in $(0, 1)$, due to Budan's theorem (Th. 3). $A_2$ is produced by applying the transformation $X \mapsto 1/(1 + X)$ to $A$, if $Var(A_2) = 0$ or $Var(A_2) = 1$ then $A$ has 0, resp. 1, real root less than 1 (Th. 1). Uspensky's [37] variant of the algorithm (see also [32]) at every step produces both polynomials $A_1$ and $A_2$. probably, as Akritas states [2], because he was unaware of Budan's theorem (Th. 3). In both variants, if the transformed polynomial has more than one sign variations, we repeat the process.

We may consider the process of the algorithm as an infinite binary tree in which the root corresponds to the original polynomial $A$. The branch from a node to a right child corresponds to the map $X \mapsto X + 1$, while to the left child to the map $X \mapsto \frac{1}{1+X}$. Notice that a sequence of $c$ transformations $X \mapsto 1 + X$ followed by one of the type $X \mapsto 1/(1 + X)$ is equivalent to two transformations, one of the type $X \mapsto c + 1/X$ followed by $X \mapsto 1 + X$. Thus Vincent's algorithm (and Uspensky's) results to a sequence of transformations like the one described in Th. 4, and so the leaves of the binary tree that we considered hold (transformed) polynomials that have no more than one sign variations, if Th. 4 holds. Akritas [1, 3] replaced a series of $X \mapsto X + 1$ transformations by $X \mapsto X + b$, where $b$ is the positive lower bound (PLB) on the positive roots of the tested polynomial. This was computed by Cauchy's bound [3, 26, 42]. This way, the number of steps is polynomial and the complexity is in $\widetilde{\mathcal{O}}_B(d^5\tau^3)$. However, it is not clear whether or how the analysis takes into account that the coefficient bitsize increases after a shift. Another issue is to bound the size of the $c_i$.

For these polynomials that have one sign variation we still have to find the interval where the real root of the initial polynomial $A$ lies. Consider a polynomial $A_n$ that corresponds to a leaf of the binary tree that has one sign variation. Notice that $A_n$ is produced after a transformation as in Th. 4, using positive integers $c_0, c_1, \ldots, c_n$. This transformation can be written in a more compact form using the convergents

$$M : X \mapsto \frac{P_n X + P_{n-1}}{Q_n X + Q_{n-1}} \tag{3}$$

where $\frac{P_{n-1}}{Q_{n-1}}$ and $\frac{P_n}{Q_n}$ are consecutive convergents of the continued fraction $[c_0, c_1, \ldots, c_n]$. Notice that (3) is a Möbius transformation, see [3, 42] for more details. Since $A_n$ has one sign variation it has one and only one real root in $(0, \infty)$, so in order to obtain the isolating interval for the corresponding real root of $A$ we evaluate the right part of Eq. (3) once over 0 and once over $\infty$. The (unordered) endpoints of the isolating interval are $\frac{P_{n-1}}{Q_{n-1}}$ and $\frac{P_n}{Q_n}$.

The pseudo-code of the CF algorithm is presented in Alg. 1. Notice that the `Interval` function orders the endpoints of the computed isolating interval and that `PLB`$(A)$ computes a lower bound on the positive roots of $A$. The initial input of the algorithm is a polynomial $A(X)$ and the trivial transformation $M(X) = X$. We need the functional $M$ in order to keep track of the transformations that we perform so that to derive the isolating intervals. Notice that Lines 1 and 1 are to be executed only when $Var(A_1) < Var(A_2)$, but in order to simplify the analysis we omit this, since it only doubles the complexity.

## 4    The complexity of the CF algorithm

The complexity of the algorithm depends on the number of transformations and the cost of each. However special care should be taken since after each transformation the bit size of the polynomial increases.

Let `disc`$(A)$ be the discriminant and `lead`$(A)$ the leading coefficient of $A$. Mahler's measure of a polynomial $A$ is $\mathcal{M}(A) = |\mathtt{lead}\,(A)| \prod_{i=1}^{d} \max\{1, |\gamma_i|\}$, where $\gamma_i$ are all the (complex) roots of $A$ [7, 42, 26, 27]. Moreover $\mathcal{M}(A) < 2^\tau \sqrt{d+1}$. We prove the following theorem, which is based on a theorem by Mignotte [26], thus extending [15, 17].

**Theorem 6** *Let $A \in \mathbb{Z}[X]$, with $\deg(A) = d$ and $\mathcal{L}(A) = \tau$. Let $\Omega$ be any set of $k$ couples of indices $(i, j)$ such that $1 \leq i < j \leq d$ and let the non-zero (complex) roots of $A$ be $0 < |\gamma_1| \leq |\gamma_2| \leq \cdots \leq |\gamma_d|$. Then*

$$2^k \mathcal{M}(A)^k \geq \prod_{(i,j) \in \Omega} |\gamma_i - \gamma_j| \geq 2^{k - \frac{d(d-1)}{2}} \mathcal{M}(A)^{1-d-k} \sqrt{\mathtt{disc}(A)}$$

**Proof.** Let $\overline{\Omega}$ be the multiset $\overline{\Omega} = \{j | (i, j) \in \Omega\}$ and $|\overline{\Omega}| = k$. We use the inequality $|a - b| \leq 2 \max\{|a|, |b|\}$ $(*)$ $a, b \in \mathbb{C}$ and the fact [26, 27] that for any root of $A$, $\frac{1}{\mathcal{M}(A)} \leq |\gamma_i| < \mathcal{M}(A)$.

In order to prove the left inequality

$$\prod_{(i,j) \in \Omega} |\gamma_i - \gamma_j| \leq 2^k \prod_{j \in \overline{\Omega}} |\gamma_j| \leq 2^k \max_{j \in \overline{\Omega}} |\gamma_j|^k \leq 2^k \mathcal{M}(A)^k.$$

Recall [7, 42, 26] that $\mathtt{disc}(A) = \mathtt{lead}\,(A)^{2d-2} \prod_{i<j} (\gamma_i - \gamma_j)^2$. For the right inequality we consider the absolute value of the discriminant of $A$:

$$\begin{aligned}
|\,\mathtt{disc}(A)| &= |\,\mathtt{lead}\,(A)|^{2d-2} \textstyle\prod_{i<j} |\gamma_i - \gamma_j|^2 \\
&= |\,\mathtt{lead}\,(A)|^{2d-2} \textstyle\prod_{(i,j)\in\Omega} |\gamma_i - \gamma_j|^2 \prod_{(i,j)\notin\Omega} |\gamma_i - \gamma_j|^2 \quad \Leftrightarrow \\
\sqrt{|\,\mathtt{disc}(A)|} &= |\,\mathtt{lead}\,(A)|^{d-1} \textstyle\prod_{(i,j)\in\Omega} |\gamma_i - \gamma_j| \prod_{(i,j)\notin\Omega} |\gamma_i - \gamma_j|
\end{aligned}$$

We consider the product $\prod_{(i,j)\notin\Omega} |\gamma_i - \gamma_j|$ and we apply $\frac{d(d-1)}{2} - k$ times inequality $(*)$, thus

$$\begin{aligned}
\textstyle\prod_{(i,j)\notin\Omega} |\gamma_i - \gamma_j| &\leq 2^{\frac{d(d-1)}{2}-k} |\gamma_1|^0 |\gamma_2|^1 \cdots |\gamma_d|^{d-1} \left(\textstyle\prod_{j\in\overline{\Omega}} |\gamma_j|\right)^{-1} \\
&\leq 2^{\frac{d(d-1)}{2}-k} \mathcal{M}(A)^{d-1} |\,\mathtt{lead}\,(A)\,|^{1-d} \mathcal{M}(A)^k
\end{aligned} \tag{4}$$

where we used the inequality $|\gamma_1|^0 |\gamma_2|^1 \cdots |\gamma_d|^{d-1} \leq |\mathcal{M}(A)/\mathtt{lead}\,(A)|^{d-1}$, and the fact [26] that, since $\forall i$, $|\gamma_i| \geq \mathcal{M}(A)^{-1}$, we have $\prod_{j\in\overline{\Omega}} |\gamma_j| \geq |\gamma_1|^k \geq \mathcal{M}(A)^{-k}$. Thus we conclude that

$$\textstyle\prod_{(i,j)\in\Omega} |\gamma_i - \gamma_j| \;\geq\; 2^{k - \frac{d(d-1)}{2}} \mathcal{M}(A)^{1-d-k} \sqrt{|\,\mathtt{disc}(A)|}$$

□

A similar theorem but with more strict hypotheses on the roots first appeared in [15] and the conditions were generalized in [17]; namely in order for the bound [15, 17] to hold the sets of indices $i$ and $j$ form an acyclic graph where each node has out-degree at most one. The bound of Th. 6 has an additional factor of $2^{d^2}$ wrt [15, 17], which plays no role when $d = \mathcal{O}(\tau)$ or when notation with $N$ is used. Moreover we loosen the hypotheses of the theorem and thus all the proofs concerning the number of steps of the subdivision-based solvers [17, 21] are dramatically simplified. Possibly a more involved proof of Th. 6 may eliminate this factor[1].

**Remark 7** *There are two simple however crucial observations about Th. 4. When the transformed polynomial has one sign variation, then the interval with endpoints $\frac{P_{n-1}}{Q_{n-1}} = [c_0, c_1, \ldots, c_{n-1}]$ and $\frac{P_n}{Q_n} = [c_0, c_1, \ldots, c_n]$ (possibly unordered) isolates a positive real root of $A$, say $\gamma_i$. Then, in order for Th. 4 to hold, it suffices to consider, instead of the separation bound $\Delta$, the quantity $|\gamma_i - \gamma_{c_i}|$, where $\gamma_{c_i}$ is the (complex) root of $A$ closest to $\gamma_i$. When the transformed polynomial has no sign variation and $[c_0, c_1, \ldots, c_n]$ is the continued fraction expansion of the (positive) real part of a complex root of $A$, say $\gamma_i$, then again it suffices to replace $\Delta$ by $|\gamma_i - \gamma_{c_i}|$.*

**Theorem 8** *The CF algorithm performs at most $\mathcal{O}(d^2 + d\tau)$ steps.*

**Proof.** Let $0 < |\gamma_1| \leq |\gamma_2| \leq \cdots \leq |\gamma_k|$, $k \leq d$ be the (complex) roots of $A$ with positive real part and let $\gamma_{c_i}$ denote the root of $A$ that is closest to $\gamma_i$.

We consider the binary tree $T$ generated during the execution of the CF algorithm. The number of steps of the CF algorithm corresponds to the number of nodes in $T$, which we denote by $\#(T)$. We use some arguments and the notation from [17] in order to prune the tree.

With each node $v$ of $T$ we associate a Möbius transformation $M_v : X \mapsto \frac{kX+l}{mX+n}$, a polynomial $A_v$ and implicitly an interval $I_v$ whose unordered endpoints can be found if we evaluate $M_v$ on $0$ and on $\infty$. Recall that $A_v$ is produced after $M_v$ is applied to $A$. The root of $T$ is associated with $A$, $M(X) = X$ (i.e $k = n = 1$, $l = m = 0$) and implicitly with the interval $(0, \infty)$.

Let a leaf $u$ of $T$ be of **type-i** if its interval $I_u$ contains $i \geq 0$ real roots. Since the algorithm terminates the leaves are of type-0 or type-1. We will prune certain leaves of $T$ so as to obtain a certain subtree $T'$ where it is easy to count the number of nodes. We remove every leaf that has a sibling that is not a leaf. Now we consider the leaves that have a sibling that is also a leaf. If both leaves are of type-1, we arbitrary prune one of them. If one of them is of type-1 then we prune the other. If both leaves are of type-0, this means that the polynomial on the parent node has at least two sign variations and thus that we are trying to isolate the (positive) real part of some complex root. We keep the leaf that contains the (positive) real part of this root. And so $\#(T) < 2 \#(T')$.

Now we consider the leaves of $T'$. All are of type-0 or type-1. In both cases they hold the positive real part of a root of $A$, the associated interval is $|I_v| \geq |\gamma_i - \gamma_{c_i}|$ (Rem. 7) and the number of nodes from a leaf to the root is $n_i$, which is such that the conditions of Rem. 5 are satisfied. Since $n_i$ is the smallest index such that the condition of Rem. 5 hold, if we reduce $n_i$ by one then the inequality does not hold. Thus

$$F_{n_i-2}|\gamma_i - \gamma_{c_i}| \leq 2d^2 \Rightarrow \phi^{n_i-3}|\gamma_i - \gamma_{c_i}| < 2d^2 \Rightarrow n_i < 4 + 2\lg d - \lg|\gamma_i - \gamma_{c_i}|$$

We sum over all $n_i$ to bound the nodes of $T'$, thus

$$\#(T') \leq \sum_{i=1}^{k} n_i \leq 2k(2+\lg d) - \sum_{i=1}^{k} \log|\gamma_i - \gamma_{c_i}| \leq 2k(2+\lg d) - \log \prod_{i=1}^{k} |\gamma_i - \gamma_{c_i}| \qquad (5)$$

In order to apply Th. 6 we should rearrange $\prod_{i=1}^{k} |\gamma_i - \gamma_{c_i}|$ so that the requirements on the indices of roots are fulfilled. This can not be achieved when symmetric products occur and thus the

worst case is when the product consists only of symmetric products i.e $\prod_{i=1}^{k/2} |(\gamma_j - \gamma_{c_j})(\gamma_{c_j} - \gamma_j)|$. Thus we consider the square of the inequality of Th. 6 taking $\frac{k}{2}$ instead of $k$ and $\mathtt{disc}(A) \geq 1$ (since $A$ is square-free), thus

$$
\begin{aligned}
\prod_{i=1}^{k} |\gamma_i - \gamma_{c_i}| &\geq \left(2^{\frac{k}{2} - \frac{d(d-1)}{2}} \mathcal{M}(A)^{1-d-\frac{k}{2}}\right)^2 \\
-\log \prod_{i=1}^{d} |\gamma_i - \gamma_{c_i}| &\leq d^2 - d - k + (2d + k - 2) \lg \mathcal{M}(A)
\end{aligned}
\tag{6}
$$

Eq. (5) becomes $\#(T') \leq 2k(2 + \lg d) + d^2 - d - k + (2d + k - 2) \lg \mathcal{M}(A)$. However for Mahler's measure it is known that $\mathcal{M}(A) \leq 2^\tau \sqrt{d+1} \Rightarrow \lg \mathcal{M}(A) \leq \tau + \lg d$, for $d \geq 2$, thus $\#(T') \leq 2k(2 + \lg d) + d^2 - d - k + (2d + k - 2)(\tau + \lg d)$. Since $\#(T) < 2\,\#(T')$ and $k \leq d$, we conclude that $\#(T) = \mathcal{O}(d^2 + d\,\tau + d \lg d)$. $\qquad\square$

### 4.1 Real root isolation

To complete the analysis of the CF algorithm we have to compute the cost of every step that the algorithm performs. In the worst case every step consists of a computation of a positive lower bound $b$ (Line 1) and three transformations, $X \mapsto b + X$, $X \mapsto 1 + X$ and $X \mapsto \frac{1}{1+X}$ (Lines 1, 1 and 1 in Alg. 1). Recall, that inversion can be performed in $\mathcal{O}(d)$. Thus the complexity is dominated by the cost of the shift operation (Line 1 in Alg. 1) if a small number of calls to $\mathtt{PLB}$ is needed in order to compute a partial quotient. We will justify this in Sec. 4.2. In order to compute this cost a bound on $\mathcal{L}(c_k) \triangleq b_k, 0 \leq k \leq m_i$ is needed, see Eq. (2).

For the analysis of the CF algorithm we will need the following:

**Theorem 9 (Fast Taylor shift)** *[41] Let $A \in \mathbb{Z}[X]$, with $\deg(A) = d$ and $\mathcal{L}(A) = \tau$ and let $a \in \mathbb{Z}$, such that $\mathcal{L}(a) = \sigma$. Then the cost of computing $B = A(a+X) \in \mathbb{Z}[X]$ is $\mathcal{O}_B(\mathtt{M}(d^2 \lg d + d^2 \sigma + d\tau))$. Moreover $\mathcal{L}(B) = \mathcal{O}(\tau + d\sigma)$.*

Initially $A$ has degree $d$ and bitsize $\tau$. Evidently the degree does not change after a shift operation. Each shift operation by a number of bitsize $b_h$ increases the bit size of the polynomial by an additive factor $d\,b_h$, in the worst case (Th. 9). At the $h$-th step of the algorithm the polynomial has bit size $\mathcal{O}(\tau + d\sum_{i=1}^{h} b_i)$ and we perform a shift operation by a number of bit size $b_{h+1}$. Th. 9 states that this can be done in $\mathcal{O}_B\left(\mathtt{M}\left(d^2 \lg d + d^2 b_{h+1} + d(\tau + d\sum_{i=1}^{h} b_i)\right)\right)$ or $\mathcal{O}_B\left(\mathtt{M}\left(d^2 \lg d + d\tau + d^2 \sum_{i=1}^{h+1} b_i\right)\right)$.

Now we have to bound $\sum_{i=1}^{h+1} b_i$. For this we use Eq. (2), which bounds $E[b_i]$. By linearity of expectation it follows that $E[\sum_{i=1}^{h+1} b_i] = \mathcal{O}(h)$ Since $h \leq \#(T) = \mathcal{O}(d^2 + d\tau)$ (Th. 8), the (expected) worst case cost of step $h$ is $\mathcal{O}_B(\mathtt{M}(d^2 \lg d + d\tau + d^2(d^2 + d\tau)))$ or $\widetilde{\mathcal{O}}_B(d^2(d^2 + d\tau))$. Finally, multiplying by the number of steps, $\#(T)$, we conclude that the overall complexity is $\widetilde{\mathcal{O}}_B(d^6 + d^5\tau + d^4\tau^2)$, or $\widetilde{\mathcal{O}}_B(d^4\tau^2)$ if $d = \mathcal{O}(\tau)$, or $\widetilde{\mathcal{O}}_B(N^6)$, where $N = \max\{d, \tau\}$.

Now let us isolate, and compute the multiplicities of, the real roots of $A_{in} \in \mathbb{Z}[X]$, which is not necessarily square-free, with $\deg(A_{in}) = d$ and $\mathcal{L}(A_{in}) = \tau$. We use the technique from [21] and compute the square-free part $A$ of $A_{in}$ using Sturm-Habicht sequences in $\widetilde{\mathcal{O}}_B(d^2\tau)$. The bit size of $A$ is $\mathcal{L}(A) = \mathcal{O}(d + \tau)$. Using the CF algorithm we isolate the positive real root of $A$ and then, by applying the map $X \mapsto -X$, we isolate the negative real roots. Finally, using the square-free factorization of $A_{in}$, which can be computed in $\widetilde{\mathcal{O}}_B(d^3\tau)$, it is possible to find the multiplicities in $\widetilde{\mathcal{O}}_B(d^3\tau)$.

The previous discussion leads to the following theorem.

**Theorem 10** *Let $A \in \mathbb{Z}[X]$ (not necessarily square-free) such that $\deg(A) = d > 2$ and $\mathcal{L}(A) = \tau$. We can isolate the real roots of $A$ and compute their multiplicities in expected time $\widetilde{\mathcal{O}}_B(d^6 + d^4\tau^2)$, or $\widetilde{\mathcal{O}}_B(N^6)$, where $N = \max\{d, \tau\}$.*

**Remark 11** *The hypothesis $d > 2$ may be replaced by $d > 4$, since real solving of polynomials of degree up to 4 can be performed in $\mathcal{O}(1)$ or $\widetilde{\mathcal{O}}_B(\tau)$ [19].*

## 4.2 Rational roots and `PLB` (Positive Lower Bound) realization

This section studies a way to compute a lower bound on the positive roots and presents its efficiency and accuracy. It seems that this is the standard approach in CF algorithms, though it is seldom discussed.

Let us consider the special case of rational roots. Their continued fraction expansion does not follow Khintchine's law, even though there are results [22] on the largest digit in such an expansion. However, recall that if $\frac{p}{q}$ is a root of $A$ then $p$ divides $a_0$ and $q$ divides $a_d$, thus in the worst case $\mathcal{L}(p/q) = \mathcal{O}(\tau)$ and so the rational roots are isolated fast. Treating them as real algebraic numbers leads to an overestimation of the number of iterations.

There is one exception to this good behavior of rational roots, namely when they are very large, well separated, and we are interested in practical complexity [4]. This is due to the fact that `PLB` must be applied many times. In [31], the authors performed a small number of Newton iterations in order to have a good approximation of a partial quotient. In [4, 5], this problem was solved by performing a homothetic transformation, $X \mapsto bX$, where $b$ is the computed bound whenever $b \geq 16$. We follow the latter approach so, after Line 1 in Alg. 1, if $b = \mathtt{PLB}(A) \geq 16$, we apply $X \mapsto bX$. The cost of applying $X \mapsto bX$ is in the worst case the same as doing a shift.

Now, let us see how $\mathtt{PLB}(A)$ is obtained in general. It is computed as the inverse of an upper bound on the roots of $X^d A(\frac{1}{X})$. In general $\mathtt{PLB}(A)$ is applied more than once in order to compute some $c_i$. However this number is very small [3, 1]. Eq. (1) implies that the probability that a partial quotient is $\leq 10$ is $\sim 0.87$, thus in general the partial quotients are of small magnitude. Moreover it is known [10, 38] that $\left| \gamma - \frac{P_n}{Q_n} \right| < \frac{1}{c_{n+1} Q_n^2}$, which means that an extremely big value of a partial quotient indicates that the *previous* approximation of the algebraic number was extremely good, thus it will need a small number of steps to isolate it.

This is how we implement `PLB`: Let $U = \{j \in \mathbb{N} \mid j < d \wedge a_j < 0\}$ and assume $a_d > 0$. Now let $C = a_d X^d + \sum_{i \in U} a_i x^j \in \mathbb{Z}[X]$. Then $C$ has exactly one nonnegative root, say $t$, which is an upper bound on the roots of $A$. We set $\mathtt{PLB}(A) = 2 \max_{a_i \in U} \left| \frac{a_i}{a_n} \right|^{1/j}$, which is nearly optimal [24], or to be more specific $t \leq \mathtt{PLB}(A) < 2t$. Actually this bound "[...] is to be recommended among all" [39], since it provides a very good approximation and is easily implementable. In our implementation we compute `PLB` only as powers of 2 so that we can take advantage of fast operations as in [33]. Notice that the cost of computing `PLB` is small, namely $\mathcal{O}(d)$ [3].

We can improve the computed bound by applying a small number of a modified Newton's iteration [35, 34] to $C$, that is guaranteed to converge rapidly. Notice that `PLB` is not a general bound on the roots, like e.g. [42, 26, 28, 27, 34], but a bound on the positive roots only, see [24, 36]. Moreover, when the number of negative coefficients is even then a bound due to Stefanescu [36] can be used which is much better.

## 5 Implementation and experiments

We have implemented the CF algorithm in SYNAPS [2] [29], which is a `C++` library for symbolic-numeric computations that provides data-structures, classes and operations for univariate and multivariate polynomials, vector, matrices, ... Our code will be included in the next public release of SYNAPS. The implementation is based on the integer arithmetic of GMP[3] (v. 4.1.4) and uses only transformations of the form $X \mapsto 2^\beta X$ and $X \mapsto X + 1$ to benefit from the fast implementations that are available in GMP. However, our implementation follows the generic programming paradigm, thus any library that provides arbitrary precision integer arithmetic can be used instead of GMP.

We restrict ourselves to square-free polynomials of degree $\in \{100, 200, \ldots, 1000\}$. Following [33], the first class of experiments concerns well-known ill-conditioned polynomials namely: Laguerre (L), first (C1) and second (C2) kind Chebyshev, and Wilkinson (W) polynomials. We also consider Mignotte (M1) polynomials $X^d - 2(101X - 1)^2$, that have 4 real roots but two of them

---

very close together, and a product $\left(X^d - 2(101X - 1)^2\right)\left(X^d - 2((101 + \frac{1}{101})X - 1)^2\right)$ of two such polynomials (M2) that has 8 real roots. Finally, we consider polynomials with random coefficients (R1), and monic polynomials with random coefficients (R2) in the range [-1000, 1000], produced by MAPLE, using 101 as a seed for the pseudo-random number generator.

We performed experiments against RS [4], which seems to be one of the fastest available software for exact real root isolation. It implements a subdivision-based algorithm using Descartes' rule of sign with several optimizations and symbolic-numeric techniques [33]. Note that we had to use RS through its MAPLE interface. Timings were reported by its internal function `rs_time()`.

We also test ABERTH [8, 9], which a numerical solver with unknown (bit) complexity but very efficient in practice, available through SYNAPS. In particular, it uses multi-precision floats and provides a floating-point approximation of all complex roots. Unfortunately, we were not always able to tune its behavior in order to produce the correct number of real roots in all the cases, i.e. to specify the input and the output precision.

In SYNAPS, there are several univariate solvers, based on Sturm sequences, Descartes' rule of sign, Bernstein basis, etc (see [18] for details and experimental results). CF is clearly faster than all these solvers, therefore we do not report on these experiments. In particular, the large inputs used here are not tractable by the Sturm-sequence solver in SYNAPS, and this is also the case for another implementation of the Sturm-sequence solver in CORE [5].

So, in Table 1, we report experiments with CF, RS, ABERTH, where the timings are in seconds. The asterisk (*) denotes that the computation did not finish after 12000s and the question-mark (?) that we were not able to tune the ABERTH solver. The experiments were performed on a 2.6 GHz Pentium with 1 GB RAM, and our code was compiled using g++ 3.3 with options -O3 -DNDEBUG.

For (M1) and (M2), there are rational numbers with a very simple continued fraction expansion that isolate the real roots which are close. These experiments are extremely hard for RS. On (M1), ABERTH is the fastest and correctly computes all real roots, but on (M2), which has 4 real roots close together, it is slower than CF. CF is advantageous on (W) since, as soon as a real root is found, transformations of the form $X \mapsto X + 1$ rapidly produce the other real roots. We were not able to tune ABERTH on (W). For (L), (C1) and (C2), CF is clearly faster than RS, while we were not able to appropriately tune ABERTH to produce the correct number of real roots. The polynomials in (R1) and (R2) have few and well separated real roots, thus the semi-numerical techniques in RS are very effective. To be more specific, RS isolates all roots using only 63 bits of accuracy (this information was extracted using the function `rs_verbose( 1)`). ABERTH is even faster on these experiments. However, even in this case, CF is faster than RS; it is a little slower than ABERTH (see Table 1).

We finally tested a univariate polynomial that appears in the Voronoi diagram of ellipses [20]. The polynomial has degree 184, coefficient bitsize 903, and 8 real roots. CF solves it in 0.12s, RS in 0.3s and ABERTH in 1.7s.

There are ways to improve our solver. First, instead of exact integer arithmetic we may use semi-numerical techniques like those in RS [33]. These techniques may be based on interval arithmetic.

**Acknowledgments** Both authors acknowledge fruitful discussions with Alkiviadis Akritas and Bernard Mourrain. The first author is grateful to Maurice Mignotte for discussions about the separation bound and to Doru Stefanescu for various discussions and suggestions about the bounds of the positive roots of polynomials. Both authors acknowledge partial support by IST Programme of the EU as a Shared-cost RTD (FET Open) Project under Contract No IST-006413-2 (ACS - Algorithms for Complex Shapes).

---

[4] fgbrs.lip6.fr/salsa/Software/index.php
[5] cs.nyu.edu/exact/core_pages/

| | | 100 | 200 | 300 | 400 | 500 | 600 | 700 | 800 | 900 | 1000 |
|---|---|---|---|---|---|---|---|---|---|---|---|
| L | CF | 0.27 | 2.24 | 9.14 | 25.27 | 55.86 | 110.13 | 214.99 | 407.09 | 774.22 | 1376.34 |
| | RS | 0.65 | 3.65 | 13.06 | 35.23 | 77.21 | 151.17 | 283.43 | 527.42 | 885.86 | 1387.45 |
| | #roots | 100 | 200 | 300 | 400 | 500 | 600 | 700 | 800 | 900 | 1000 |
| C1 | CF | 0.11 | 0.85 | 3.16 | 8.61 | 19.67 | 38.23 | 77.75 | 139.18 | 247.11 | 414.51 |
| | RS | 0.21 | 1.36 | 3.80 | 10.02 | 23.15 | 46.02 | 82.01 | 150.01 | 269.35 | 458.67 |
| | #roots | 100 | 200 | 300 | 400 | 500 | 600 | 700 | 800 | 900 | 1000 |
| C2 | CF | 0.11 | 0.77 | 3.14 | 8.20 | 19.28 | 38.58 | 73.59 | 133.52 | 233.48 | 386.61 |
| | RS | 0.23 | 1.48 | 3.80 | 9.84 | 23.28 | 46.34 | 83.58 | 146.04 | 273.00 | 452.77 |
| | #roots | 100 | 200 | 300 | 400 | 500 | 600 | 700 | 800 | 900 | 1000 |
| W | CF | 0.11 | 0.76 | 2.54 | 6.09 | 12.07 | 21.43 | 34.52 | 53.35 | 81.88 | 120.21 |
| | RS | 0.09 | 0.59 | 2.25 | 6.34 | 14.62 | 29.82 | 55.47 | 104.56 | 179.23 | 298.45 |
| | #roots | 100 | 200 | 300 | 400 | 500 | 600 | 700 | 800 | 900 | 1000 |
| M1 | CF | 0.02 | 0.08 | 0.21 | 0.42 | 0.73 | 1.19 | 1.84 | 2.75 | 4.16 | 6.22 |
| | RS | 7.83 | 287.27 | 1936.48 | 7328.86 | * | * | * | * | * | * |
| | ABERTH | 0.01 | 0.04 | 0.07 | 0.11 | 0.12 | 0.26 | 0.43 | 0.37 | 0.47 | 0.90 |
| | #roots | 4 | 4 | 4 | 4 | 4 | 4 | 4 | 4 | 4 | 4 |
| M2 | CF | 0.08 | 0.43 | 1.10 | 2.78 | 4.71 | 8.67 | 18.26 | 25.28 | 40.15 | 60.10 |
| | RS | 1.24 | 144.64 | 1036.785 | 4278.275 | 12743.79 | * | * | * | * | * |
| | ABERTH | 0.04 | 0.78 | 3.24 | ? | ? | ? | ? | ? | ? | ? |
| | #roots | 8 | 8 | 8 | 8 | 8 | 8 | 8 | 8 | 8 | 8 |
| R1 | CF | 0.001 | 0.04 | 0.07 | 0.33 | 0.06 | 0.37 | 0.66 | 0.76 | 1.03 | 1.77 |
| | RS | 0.026 | 0.09 | 0.11 | 0.68 | 0.22 | 0.89 | 0.95 | 0.69 | 1.55 | 2.09 |
| | ABERTH | 0.02 | 0.03 | 0.07 | 0.14 | 0.21 | 0.31 | 0.44 | 0.51 | 0.64 | 0.80 |
| | #roots | 4 | 4 | 2 | 6 | 2 | 4 | 4 | 2 | 4 | 4 |
| R2 | CF | 0.01 | 0.04 | 0.08 | 0.36 | 0.14 | 0.38 | 0.74 | 0.77 | 1.24 | 1.42 |
| | RS | 0.05 | 0.23 | 0.47 | 1.18 | 0.81 | 1.64 | 2.68 | 3.02 | 4.02 | 4.88 |
| | ABERTH | 0.01 | 0.05 | 0.08 | 0.14 | 0.23 | 0.33 | 0.44 | 0.55 | 0.67 | 0.83 |
| | #roots | 4 | 4 | 4 | 6 | 4 | 4 | 6 | 4 | 6 | 4 |

Table 1: Experimental results